\documentclass{article}

\usepackage{microtype}
\usepackage{graphicx}
\usepackage{subcaption}
\usepackage{booktabs} 
\usepackage{natbib}
\usepackage{xcolor}
\usepackage{commath}
\usepackage{xcolor}
\usepackage{tabularx,lipsum}
\usepackage{booktabs}
\usepackage{makecell}
\usepackage{hyperref}

\usepackage[accepted]{icml2020}

\icmltitlerunning{Reconstruction of Total Solar Irradiance by Deep Learning}
\newcolumntype{R}{>{\raggedright\arraybackslash}X}
\newcolumntype{W}{>{\centering\arraybackslash}X}
\begin{document}

\twocolumn[
\icmltitle{Reconstruction of Total Solar Irradiance by Deep Learning}


\begin{icmlauthorlist}
\icmlauthor{Yasser Abduallah}{goo}
\icmlauthor{Jason T. L. Wang}{goo}
\icmlauthor{Yucong Shen}{goo}
\icmlauthor{Khalid A. Alobaid}{goo}
\icmlauthor{Serena Criscuoli}{ed}
\icmlauthor{Haimin Wang}{to}

\end{icmlauthorlist}

\icmlaffiliation{to}{Institute for Space Weather Sciences, 
New Jersey Institute of Technology, University Heights, Newark, NJ 07102, USA}
\icmlaffiliation{goo}{Department of Computer Science, New Jersey Institute of Technology, Newark, NJ 07102, USA}
\icmlaffiliation{ed}{National Solar Observatory, 3665 Discovery Dr., Boulder, CO 80303, USA}

\icmlcorrespondingauthor{Jason T. L. Wang}{wangj@njit.edu}

\icmlkeywords{Machine Learning}

\vskip 0.3in
]



\printAffiliationsAndNotice{\icmlEqualContribution} 

\begin{abstract}
The Earth's primary source of energy is the radiant energy generated by the Sun, which is referred to as solar irradiance,
or total solar irradiance (TSI) when all of the radiation is measured.
A minor change in the solar irradiance can have a significant impact on the Earth's climate and atmosphere.
As a result, studying and measuring solar irradiance is crucial in understanding climate changes and solar variability.
Several methods have been developed to reconstruct total solar irradiance for long and short periods of time;
however, they are physics-based and rely on the availability of data, which does not go beyond 9,000 years.
In this paper we propose a new method, called TSInet, to reconstruct total solar irradiance by deep learning
for short and long periods of time that span beyond the physical models' data availability.
On the data that are available, our method agrees well with the state-of-the-art physics-based reconstruction models.
To our knowledge, this is the first time that deep learning has been used to reconstruct total solar irradiance
for more than 9,000 years.
\end{abstract}

\section{Introduction}\label{sec:introduction}
Solar irradiance is the primary source of energy for our Earth 
\citep{WhereSunGetEnergyKrenEaAll2017}, 
and is a key input for climate models and changes
\citep{SunRoleInClimateHansen2000, SIVariabilityCompareSORCE2011, NewLowerValueOfTSIKopp2011, 
ClimateForcingReconstructions2011, SIVariabilityClimateSolanki2013}. 
It is described in terms of total solar irradiance (TSI) when all of the radiation is measured.
Irradiance is defined as the amount of light energy from an object that is hitting a square meter of another object each second. 
The solar irradiance is the amount of light energy from the Sun's entire disk measured at the Earth, 
and it is known to vary over different temporal scales, in a manner that is strongly wavelength dependent
\citep{ kopp2016}.
TSI variability affects the Earth's
atmosphere and climate in many ways 
\citep{SolarInfluenceOnClimateGray2010}. 
To understand the effect of solar radiation on our Earth's climate changes, 
solar irradiance records for long periods of time 
are required 
\citep{jungclaus2017}. 
Since systematic measurements of irradiance started only in the late seventies, many models were introduced to provide
irradiance records dating back to times ranging from century to millennia. 
All such models are based on the empirical evidence that irradiance variability is modulated by surface magnetism 
\citep{domingo2009}, 
while the  approaches adopted in the different models are  mostly driven by 
the type of proxies of the magnetic field available at the temporal scales considered. 

Most of the published models aim to reconstruct irradiance variability up to a few centuries into the past
\citep{EvolutionOfSunSSIMaunderMinLean2000,ReconstructionofSTISince1700Krivova2007}. 
Such models are intended to address the impact 
of solar variability on Earth's increase of temperatures registered from the pre-industrial era, 
and mostly make use of sunspot, or sunspot-group number, as proxy of the surface magnetic activity. 
A few models have instead been proposed in the literature aiming at reconstructing irradiance variations at longer temporal scales. 
Because at those times telescopic observations were not available, 
such reconstructions necessarily make use of indirect proxies. 
These mostly consist of radioisotopes like  ${^{14}}$C, $^{10}$Be and 
nitrate-related species \citep{usoskin2017}, 
which are generated by the  interaction
of energetic particles with the Earth's atmosphere, whose flux, in turn, is regulated by the heliospheric magnetic field.

Some of the historical irradiance reconstruction models used linear regression relationships between the irradiance measured at modern times
and input proxies.
More complex techniques make use of geomagnetic models to estimate from radioisotopes
the \textit{open} and \textit{closed} components of the solar magnetic field, 
from which the distribution of magnetic features over the Sun's disk is recovered.   
The most recent state-of-the-art model of this kind was developed by \citet{Wu9000YearSIReconstruction}, 
who reconstructed TSI for the previous 9 millennia, 
making use of two different cosmogenic isotopes, ${^{10}}$Be and ${^{14}}$C, derived from various datasets 
\citep{WuSolarActivitySunSpotReconstruction2018}. 
All of the published models reconstructed solar irradiance based on physics properties
\citep{Wu9000YearSIReconstruction, EffectOfUVCriscuoli_2019}. 
		 
In this paper, we present the first deep learning model, called TSInet,
to reconstruct total solar irradiance for more than 9,000 years. 
We reconstruct the entire 9,000 years already covered in the recent reconstruction 
by \citet{Wu9000YearSIReconstruction} and an 
additional 1,000 years when physical data are not available. 
Our deep learning model does not rely on proxies; 
therefore our model is not affected by uncertainties in the proxies 
including errors in their measurements and estimates. 
Furthermore, our model can be extended back at times when proxies are not available.
When the data are available, TSInet agrees well with 
the state-of-the-art physics-based reconstruction models on the available data.   

The rest of this paper is organized as follows.	
Section \ref{sec:data} describes the datasets used for
training and testing TSInet.
Section \ref{sec:methodology} details the architecture and algorithms employed by TSInet.
Section \ref{sec:results} reports experimental results.
Section \ref{sec:conclusions} concludes the paper and points out some directions for future research.

\section{Data}\label{sec:data}
\begin{figure} [h]
	\centering
	\includegraphics[width=\columnwidth]{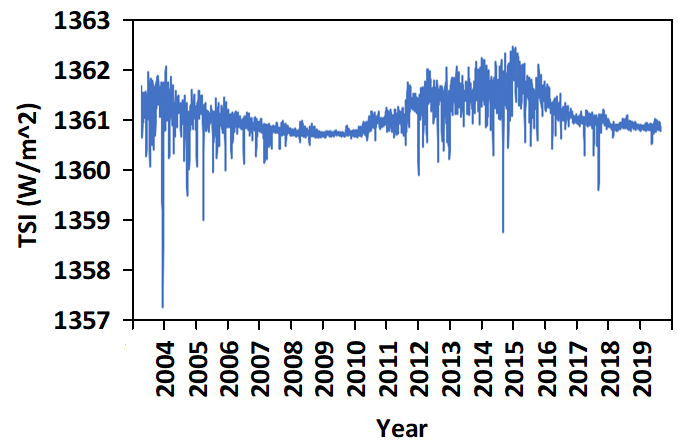}	
	\caption{SORCE total solar irradiance (TSI) data from 2003 to present.}
	\label{SORCEdata}
\end{figure}

In this work we use measurements of the TSI provided by the 
Total Irradiance Monitor 
aboard the SOlar Radiation and Climate Experiment \citep[SORCE;][]{rottman2005}
and available at \url{http://lasp.colorado.edu/home/sorce/data/tsi-data/}. 
This dataset, used as our training set, contains daily TSI measurements carried out from 2003 to present.
Figure \ref{SORCEdata} illustrates the SORCE time series dataset showing
the total solar irradiance over time. 

Our testing set contains measurements from TCTE Total Solar Irradiance daily averages,
available at \url{http://lasp.colorado.edu/lisird/data/tcte_tsi_24hr/}. 
Total Solar Irradiance Calibration Transfer Experiment (TCTE) (\url{http://lasp.colorado.edu/home/missions-projects/quick-facts-tcte/}) 
measurements are made by the LASP TCTE Total Irradiance Monitor (TIM) instrument aboard the U.S. Air Force's STPSat-3 spacecraft. 
This TIM has been measuring total solar irradiance since late 2013. 

In addition, we adopt the following publicly available datasets obtained, over different temporal ranges, 
by different physics-based models, which will be used as testing sets in our work.
\begin{itemize}		
\item {\textbf{NRLTSI2 Daily Averages}, available at \url{http://lasp.colorado.edu/lisird/}, 
is the daily climate record of total solar irradiance from 1882 to present. 
It is constructed using 
version 2 
of the Naval Research Laboratory's (NRL) solar variability model (NRLTSI2). 
The NRLTSI2 model computes TSI based on the changes of 
the quiet Sun conditions arising from bright faculae and dark sunspots on the solar disk. 
It uses linear regression between proxies of solar sunspot and facular features, as well as irradiance 
observations from SORCE.}
\item{\textbf{SATIRE-S} (Spectral And Total Irradiance
REconstruction model - Space era), available at \url{http://www2.mps.mpg.de/projects/sun-climate/data.html},
provides daily reconstruction of solar irradiance in the period of 1974 -- 2013. 
Irradiance is reconstructed by combining the area coverage of magnetic and quiet features as derived by 
full-disk magnetograms and continuum images of the Sun, 
together with spectral syntheses obtained by one-dimensional, static, atmosphere models 
\citep{SolarReconstruction1974-2013YeoEtAl2014}.}	
\item{\textbf{SATIRE-M} (Spectral And Total Irradiance
REconstruction model - Millennia), available at \url{http://www2.mps.mpg.de/projects/sun-climate/data.html},
is similar to SATIRE-S, but the area coverage of magnetic structures is estimated by making use of a model 
which relies on the sunspot number as derived by indirect proxies
\citep{WuSolarActivitySunSpotReconstruction2018}. 
This model provides decennial averages and
reconstructs the solar irradiance over the last 9,000 years. 
The model is used to reconstruct decadal total TSI.}	
\end{itemize} 

\begin{figure} 
	 	\centering
	 	\includegraphics[width=\columnwidth]{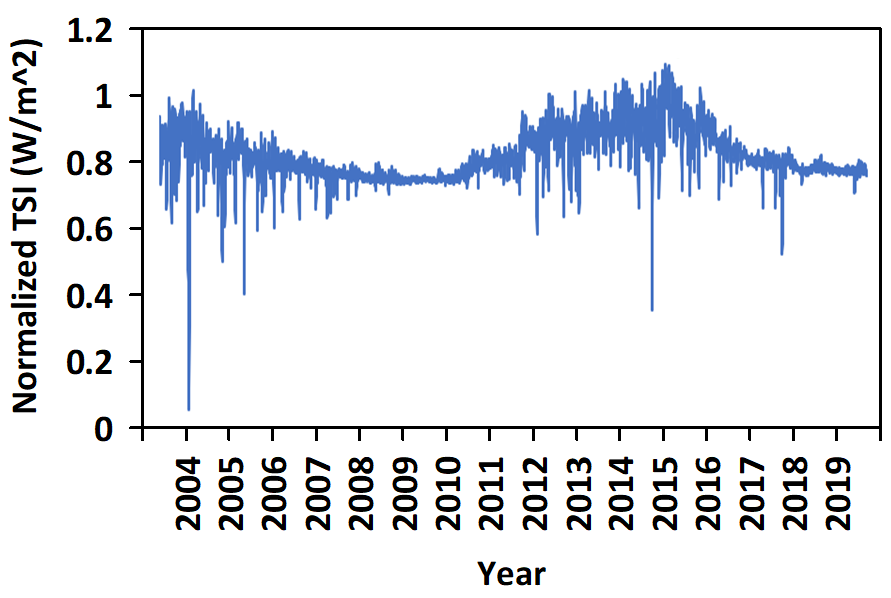}
	 	\caption{SORCE total solar irradiance (TSI) data from 2003 to present scaled by the min-max normalization technique.}
	 	\label{fig:datascaling}
	 \end{figure}

The total solar irradiance values range from 1356.656 to 1363.525. 
We use a feature scaling technique, also known as data normalization, to normalize the range of data to increase the cohesion of the TSI values. 
Specifically, we use the min-max normalization that is calculated as follows:
	\begin{equation}
		\hat{v_i} = \frac{v_i - \text{min}(S)}{\text{max}(S) - \text{min}(S)}
	\end{equation} 
where $\hat{v_i}$ (${v_i}$, respectively) is the normalized value (actual value, respectively) at time point $i$ and $S$ represents the input data set.
Fig.~\ref{fig:datascaling} shows the SORCE TSI time-series scaled by the min-max normalization technique.
 
\section{Proposed Method}\label{sec:methodology}

\begin{figure*}[h]
		\centering
		\includegraphics[width=1.3\columnwidth]{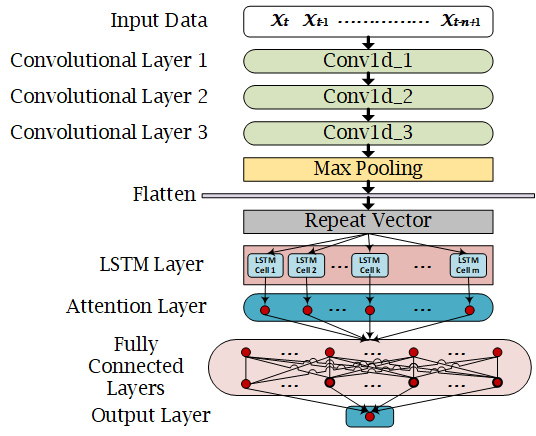}
		\caption{Architecture of TSInet.}
		\label{fig:contextualarchitecture}
\end{figure*}

\subsection{Architecture and Training of TSInet} 
\label{TSInet-training}
Figure \ref{fig:contextualarchitecture} presents the architecture of our TSInet.
Let $t$ be the latest time point.
Data sample $x_{t}$ contains $w$ values $v_{t}, v_{t-1}, \ldots , v_{t-w+1}$ and the label $v_{t-w}$ where
$v_{t}$ is the TSI value at time point $t$.
(In the study presented here, the time window, $w$, is set to 7.)
We train TSInet with multiple batches.
In the first batch, we use the $n$ training data samples, 
$x_{t}, x_{t-1}, \ldots , x_{t-n+1}$, to train TSInet.
(In the study presented here, the number of input data samples, $n$, is set to 10.)
The label of the $n$ training data samples is determined by the label of the last data sample (i.e., $x_{t-n+1}$).
In the second batch, we use the next $n$ training data samples,
$x_{t-n}, x_{t-n-1}, \ldots , x_{t-2n+1}$, to train TSInet.
The label of the $n$ training data samples is determined by the label of $x_{t-2n+1}$.
In the third batch, we use the following $n$ training data samples,
$x_{t-2n}, x_{t-2n-1}, \ldots ,  x_{t-3n+1}$, to train TSInet.
The label of the $n$ training data samples is determined by the label of $x_{t-3n+1}$.
We continue this training process until all TSI values in the training set are used.
For every two adjacent data samples $x_{i}$, $x_{i-1}$,
they overlap on $w-1$ TSI values, namely $v_{i-1}, v_{i-2}, \ldots , v_{i-w+1}$.

Our TSInet consists of three convolutional layers (Conv1d\_1, Conv1d\_2, and Conv1d\_3)
where the kernel slides along 1 dimension on the time series, 
a max pooling layer, a flatten layer, 
a repeat vector layer, 
an LSTM (long short-term memory) layer, an attention layer, two fully connected layers, and an output layer.
The convolutional layers Conv1d\_1, Conv1d\_2, and Conv1d\_3 are configured with 64, 128, and 256 kernels, respectively. 
The kernel sizes of the convolutional layers are 
(3 $\times$ 1 $\times$ 64), 
(3 $\times$ 1 $\times$ 128), and 
(3 $\times$ 1 $\times$ 256), respectively.  
The max-pooling layer is configured with a pooling factor of 2. 
The output from the three convolutional layers is flattened by the flatten layer
 and transformed into a sequence, 
also known as a feature vector. 
The repeat vector layer repeats the feature vector to reshape and prepare it as the input to the LSTM layer.
The LSTM layer in our architecture contains $m$ LSTM cells (in this study, $m$ is set to 10). 
The attention layer with $m$ neurons is used to focus on the relevant information in each time step as done in \citep{AttentionLayerBahdanau2015}.
Each of the two fully connected layers has 200 neurons.
The activation function used in our model is ReLU (rectified linear unit).
TSInet produces as output a predicted TSI value.

The proposed TSInet is implemented in Python, Keras, and Tensorflow. 
We use adaptive moment estimation \citep[Adam;][]{DeepLearningBook2LeCun2015,DeepLearningBookDBLP:books/daglib/0040158} 
as the network optimizer, which is a stochastic gradient descent algorithm that can update network weights 
based on training data. 
Adam is configured with a learning rate of 0.003 and a weight decay of 0.000005
 to regularize the weights 
and minimize the test error during training in each epoch. 
Other Adam parameters ($\beta_1$, $\beta_2$, respectively) are set to default values (0.9, 0.999, respectively).
To achieve faster back-propagation convergence, 
we adopt the mini-batch strategy described in \citep{DeepLearningBook2LeCun2015,DeepLearningBookDBLP:books/daglib/0040158}. 
The number of epochs is set to 10 by default.

\subsection{Reconstruction of Total Solar Irradiance}\label{sec:modelimplementation}
After describing the architecture and training procedure of TSInet,
we now turn to the algorithms for reconstructing total solar irradiance (TSI) in a testing set.
We develop two reconstruction algorithms: (i) single-step or 1-step reconstruction; 
(ii) multi-step or $k$-step, $k > 1$,  reconstruction.

Let $t$ be the latest time point.
With single-step reconstruction, we begin by considering the $n$ testing data samples
$x_{t}, x_{t-1}, \ldots , x_{t-n+1}$ where 
$x_{t}$ contains the $w+1$ TSI values
$v_{t}, v_{t-1}, \ldots , v_{t-w+1}, v_{t-w}$ in the testing set.
Our TSInet model, which is trained as described in Section \ref{TSInet-training}, takes as input the 
$n$ testing data samples and
predicts the label of the last testing data sample (i.e., $x_{t-n+1}$),
which is treated as the label of the $n$ testing data samples.
We then use the $n$ testing data samples together with the predicted label 
to re-fit or re-train TSInet.
The re-trained TSInet then takes as input the next $n$ testing data samples
$x_{t-1}, x_{t-2}, \ldots , x_{t-n}$ and
predicts the label of the last testing data sample (i.e., $x_{t-n}$),
which is treated as the label of the $n$ testing data samples.
We again use the $n$ testing data samples together with the predicted label 
to re-fit or re-train TSInet.
The re-trained TSInet then takes as input the following $n$ testing data samples
$x_{t-2}, x_{t-3}, \ldots , x_{t-n-1}$ and 
predicts the label of the last testing data sample (i.e., $x_{t-n-1}$),
which is treated as the label of the $n$ testing data samples.
We then use the $n$ testing data samples together with the predicted label to re-train TSInet.
We continue this predicting-retraining process until all TSI values (labels) in the testing set have been predicted,
at which point we have reconstructed all the TSI values in the testing set.

With multi-step reconstruction, we begin by considering the first batch containing the $n$ testing data samples
$x_{t}, x_{t-1}, \ldots , x_{t-n+1}$ in the testing set.
Our trained TSInet takes as input these $n$ testing data samples and
predicts the label of the last testing data sample (i.e., $x_{t-n+1}$),
which is treated as the label of the $n$ testing data samples.
Then, the same TSInet model takes as input the second batch containing the next $n$ testing data samples 
$x_{t-1}, x_{t-2}, \ldots , x_{t-n}$ and
predicts the label of the last testing data sample (i.e., $x_{t-n}$),
which is treated as the label of the $n$ testing data samples.
We keep on using the same TSInet model until the model takes as input the $k$th batch containing the
$n$ testing data samples
$x_{t-k+1}, x_{t-k}, \ldots , x_{t-n-k+2}$ and 
predicts the label of the last testing data sample (i.e., $x_{t-n-k+2}$),
which is treated as the label of the $n$ testing data samples.
We then use the $k$ batches, where each batch contains $n$ testing data samples together with their predicted label, 
to retrain our TSInet as shown in Figure \ref{fig:contextualarchitecture}.
Then we use the re-trained model to predict the labels for the next $k \times n$ testing data samples.

The difference between single-step reconstruction and multi-step reconstruction is that 
the former retrains TSInet once using one batch containing $n$ testing data sample in every one step 
while the latter retrains TSInet once using $k$ batches containing $k \times n$ testing data samples in every $k$ steps.

\section{Experiments and Results}\label{sec:results}
	\subsection{Performance Metrics}
We conducted a series of experiments to evaluate the performance of the proposed TSInet and
compare it with related methods.		
The performance metrics used here are the root mean square error (RMSE)
and Pearson correlation coefficient (CORR).
RMSE is calculated as follows:
\begin{equation}\label{eq:rmse}
\text{RMSE} = \sqrt{\frac{1}{n}\sum_{i=1}^{n}(\hat{y_i} - {y_i})^2}
\end{equation}
where $\hat{y_i}$ ($y_i$, respectively) represents the predicted TSI value (actual TSI value, repectively) at time point $i$.
RMSE 
measures the differences between the actual TSI
values and the predicted TSI values by a method.
The lower the RMSE value, the more accurate the method is. 
CORR is calculated as follows:
\begin{equation}\label{eq:corr}
		\text{CORR} = \frac{\sum_{i}(\hat{y_i} - \mu({\hat{y}}))(y_i - \mu(y))}
					{\sqrt{\sum_{i}^{\textcolor{white}{2}}(\hat{y_i} - \mu(\hat{y}))^2} \text{ }\sqrt{\sum_{i}^{\textcolor{white}{2}}({y_i} - \mu({y}))^2}}
\end{equation}
where 
$\mu({\hat{y}})$ denotes the mean of all predicted TSI values and $\mu(y)$ denotes the mean of all actual TSI values.
CORR is used to measure how strong the relationship between the predicted and actual TSI
values is. 
CORR ranges from 1 to -1, 
where 1 means there is a strong positive correlation, 
-1 means there is a strong negative correlation, 
and 0 means there is no correlation between the predicted and actual TSI values.

\subsection{Single-Step vs. Multi-Step Reconstruction Algorithms}
\begin{figure}[h]
	\centering
	\includegraphics[width=\columnwidth,height=155px]{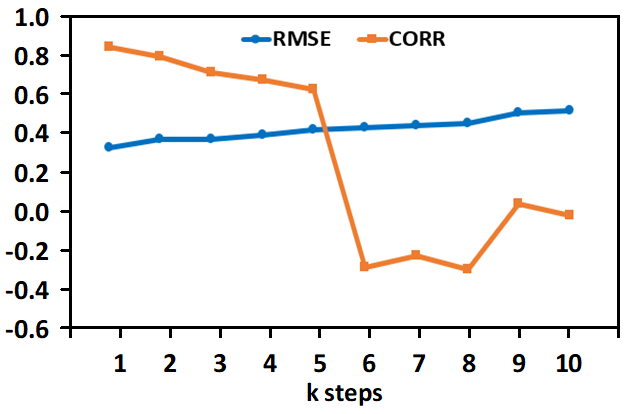}
	\caption{RMSE and CORR values obtained by TSInet for $k$-step, $k = 1, 2, \ldots, 10$, reconstruction of TSI.}  
	\label{fig:nstespvsrmsecorr}
\end{figure}

\begin{figure}[h]
	\centering
	\includegraphics[width=\columnwidth,height=155px]{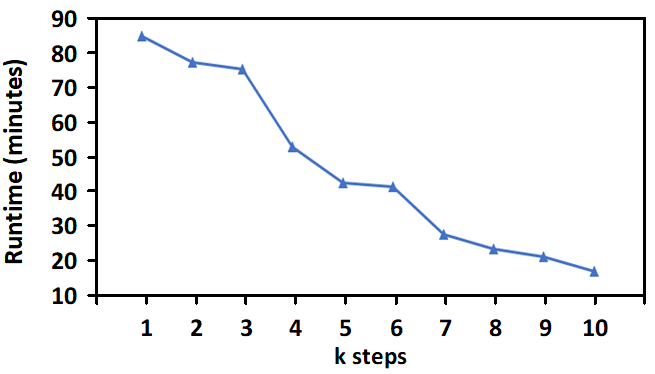}
	\caption{Runtime needed by TSInet for $k$-step, $k = 1, 2, \ldots, 10$, reconstruction of TSI.}  
	\label{fig:nstespvstime}
\end{figure}

In this experiment we compared the single-step (i.e., 1-step) and multi-step (i.e.,  $k$-step) reconstruction algorithms 
described in Section \ref{sec:modelimplementation}.
We used the SORCE training set to train TSInet and reconstructed the TSI values in the TCTE testing set
for varying $k$, $k = 1, \ldots , 10$.
For each $k$, we computed the performance metrics and recorded the runtime used by the algorithms.
\figref{fig:nstespvsrmsecorr} shows the performance metrics, RMSE and CORR, for varying $k$.
it can be seen from the figure that the performance of TSInet degrades as $k$ increases.
This happens because the TSInet model is refitted more often, and hence is more accurate, when $k$ is smaller.
On the other hand, smaller $k$ requires more runtime, as shown in \figref{fig:nstespvstime}.
In subsequent experiments, we fixed $k = 5$ as it achieved good performance while requiring reasonable runtime.

\subsection{Ablation Tests}
In this experiment we performed ablation tests to analyze and evaluate the components of our TSInet framework 
by considering two models based on TSInet: CNN and LSTM. 
The CNN model is a subnet of TSInet, keeping the three convolutional layers, max pooling layer, flatten layer, repeat vector layer, attention layer, 
two fully connected layers and output layer, but removing the LSTM layer.
The LSTM model is also a subnet of TSInet, keeping the LSTM layer, attention layer, two fully connected layers and output layer, 
but removing the three convolutional layers, max pooling layer, flatten layer, and repeat vector layer. 

\begin{figure}[h]
	\centering
	\includegraphics[width=\columnwidth]{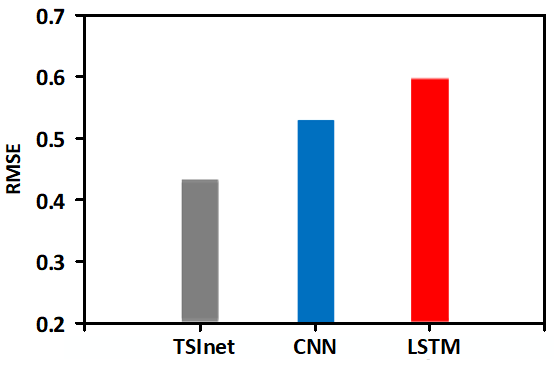}
	\caption{Comparison of RMSE values from TSInet, CNN and LSTM.}
	\label{fig:nstespablationrmse}
\end{figure}

\begin{figure}[h]
	\centering
	\includegraphics[width=\columnwidth]{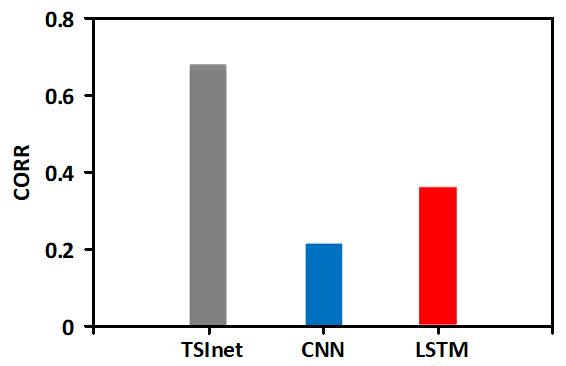}
	\caption{Comparison of CORR values from TSInet, CNN and LSTM.}
	\label{fig:nstespablationcorr}
\end{figure}

\figref{fig:nstespablationrmse} (\figref{fig:nstespablationcorr}, respectively) presents the RMSE (CORR, respectively) results 
from TSInet, CNN and LSTM.
It can be seen from the figures that TSInet yields the best accuracy and correlation among the three methods.
This happens because CNN learns characteristics from the input data but it lacks temporal components to deeply analyze the time series information in the data.
On the other hand, LSTM analyzes the temporal correlation between the input and output data, 
but it works on the raw input data without learning additional characteristics to strengthen the correlation between the data entries. 
TSInet combines the characteristics it learns in the CNN network and temporal correlation it learns in the LSTM network. 
Therefore, TSInet achieves the best performance.

\subsection{Comparison with Related Methods}
\begin{figure}[h]
	\centering
	\includegraphics[width=\columnwidth,height=155px]{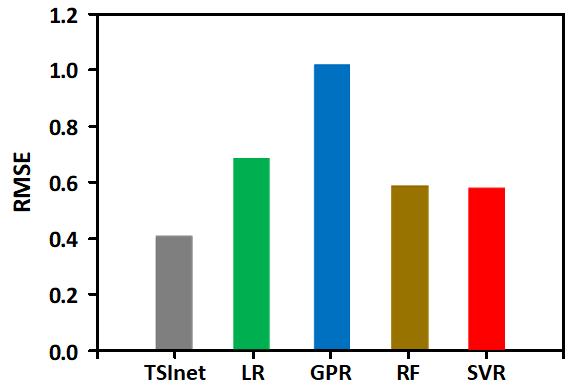}
	\caption{Comparison of RMSE values from five TSI reconstruction methods on the TCTE dataset.} 
	\label{fig:rmsealgorithmstcte}
\end{figure}

\begin{figure}[h]
	\centering
	\includegraphics[width=\columnwidth,height=155px]{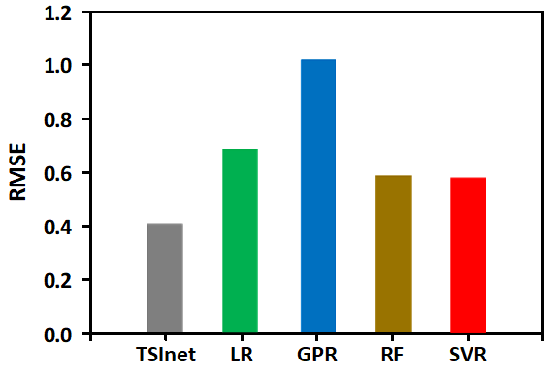}
	\caption{Comparison of RMSE values from five TSI reconstruction methods on the NRLTSI2 dataset.}
	\label{fig:rmsealgorithmsnrltsi2}
\end{figure}

\begin{figure}[h]
	\centering
	\includegraphics[width=\columnwidth,height=155px]{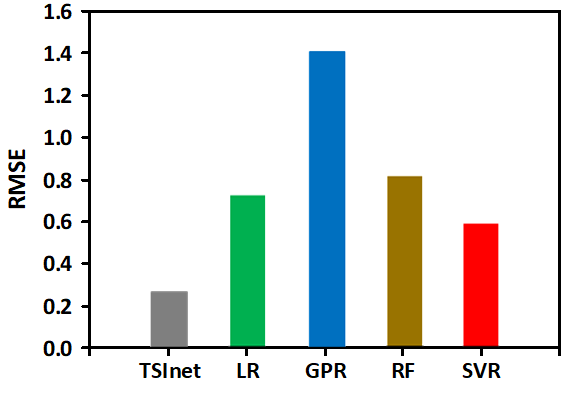}
	\caption{Comparison of RMSE values from five TSI reconstruction methods on the SATIRE-S dataset.}
	\label{fig:rmsealgorithmssatires}
\end{figure}

In this experiment we compared TSInet with four closely related machine learning algorithms including
linear regression \citep[LR;][]{LR2005SongEaAl}, 
Gaussian process regression \citep[GPR;][]{GPRSpationTempRansalu,GaussianGPalshedivat2017srk}, 
random forest \citep[RF;][]{Liu..Wang..Solar..2017ApJ...843..104L}, and 
support vector regression \citep[SVR;][]{BobraCME2016ApJ...821..127B}. 
Figures \ref{fig:rmsealgorithmstcte}, \ref{fig:rmsealgorithmsnrltsi2}, and \ref{fig:rmsealgorithmssatires} 
present the RMSE results for the five methods on the 
TCTE, NRLTSI2, and SATIRE-S datasets respectively. 
The figures show that TSInet achieves the best performance among the five methods in terms of RMSE.
The CORR results are similar and omitted here.

To assess whether the results obtained by our TSInet agree with entries in the testing datasets,
we performed the Wilcoxon signed-rank test \citep{WilcoxonPValue,WilcoxonPValue1959}.
According to the test,
the difference between the TSInet results and entries in TCTE (NRLTSI2, SATIRE-S respectively)
is not statistically significant with $p$ = 0.01552 $<$ 0.05 (0.0001 $<$ 0.05, 0.0001 $<$ 0.05 respectively).

\subsection{Reconstruction of TSI on the SATIRE-M Dataset}
SATIRE-M contains decennial averages and
is comprised of solar irradiance over the last 9,000 years. 
Each entry in the SATIRE-M dataset represents an average of 10 years.
However, our TSInet reconstructs daily TSI. 
To reconstruct solar irradiance on the SATIRE-M dataset, we employ the following technique.
Recall that the SATIRE-S dataset provides daily reconstruction of solar irradiance in the period of 1974 -- 2013. 
We first use TSInet to reconstruct total solar irradiance beyond 1974 on SATIRE-S.
Then we take 10-year averages on the reconstructed TSI values.

\begin{figure}
	\centering
	\includegraphics[width=1\columnwidth]{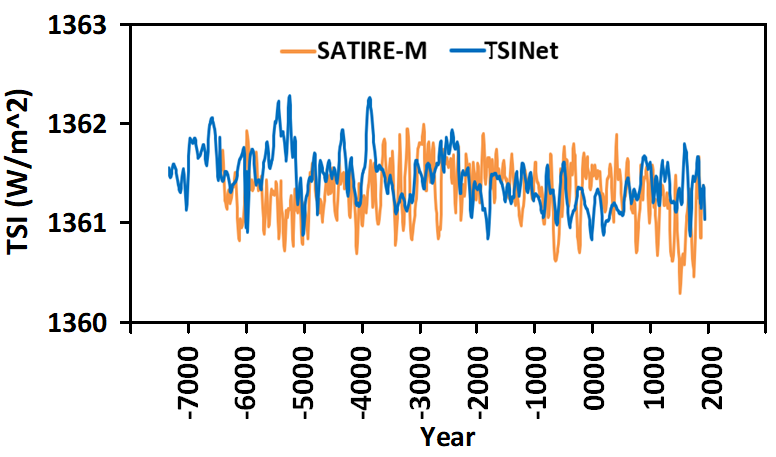}
	\caption{SATIRE-M reconstruction using TSInet plus reconstruction of solar irradiance for additional1,000 years.}
	\label{fig:satiremtsinet}
\end{figure}

\figref{fig:satiremtsinet} compares the 10-year averages obtained by TSInet with the entries in SATIRE-M.
TSInet's results agree mostly with entries in SATIRE-M.
According to the Wilcoxon signed-rank test \citep{WilcoxonPValue,WilcoxonPValue1959},
the difference between TSInet's results and SATIRE-M entries is not statistically significant ($p$ = 0.000297 $<$ 0.05).
\figref{fig:satiremtsinet} also shows that our TSInet model is capable of reconstructing total solar irradiance beyond 9,000 years. 
We reconstructed total solar irradiance for additional 1,000 years beyond the SATIRE-M data. 

\section{Conclusions and Future Work} \label{sec:conclusions}
The Earth's primary source of energy is the radiant energy from the Sun. 
This energy is known as solar irradiance, or total solar irradiance (TSI) when all of the radiation is measured.
The changes in solar irradiance has a significant impact on Earths' atmosphere and climate. 
Therefore, studying and reconstructing solar irradiance is crucial in solar physics. 
Existing methods for solar irradiance reconstruction are all based on physics-based models
\citep{EMPIRESIYeo2017EMPIREAR, Wu9000YearSIReconstruction}.
In this paper we presented the first deep learning method (TSInet) for reconstructing 
total solar irradiance (TSI).
Experimental results showed that results from our TSInet agree well with those from the physics-based models.
When compared to closely related machine learning methods, 
TSInet achieves the best performance among the methods.
TSInet does not reply on physics properties such as proxies, and hence it can be extended back at times when proxies were not available.
We demonstrated here that TSInet is able to reconstruct TSI for more than 9 millennia.
In future work we plan to extend TSInet to analyze other time series data that arise in solar physics.

\bibliography{ya-tsinet}
\bibliographystyle{icml2020}

\end{document}